\documentclass[10pt,a4paper,twoside]{article}
\usepackage{epsfig}
\usepackage{baltlat6}
\usepackage{wrapfig}
\usepackage{lscape}
\begin{document}
\ \ \vspace{-0.5mm}

\setcounter{page}{421}
\vspace{-2mm}

\titlehead{Baltic Astronomy, vol.\,16, 421--429, 2007}

\titleb{ACCURACY OF STAR CLUSTER PARAMETERS FROM\\ INTEGRATED
{\it UBVRI} PHOTOMETRY}

\begin{authorl}
\authorb{D.~Narbutis,}{}
\authorb{A.~Brid\v{z}ius,}{}
\authorb{R.~Stonkut\.{e}}{} and
\authorb{V.~Vansevi\v{c}ius}{}
\end{authorl}

\moveright-3.2mm \vbox{
\begin{addressl}
\addressb{}{Institute of Physics, Savanori\c{u} 231, Vilnius LT-02300, Lithuania}
\end{addressl}}

\submitb{Received 2007 November 30; accepted 2007 December 14}

\begin{summary} We study the capability of the {\it UBVRI} photometric
system to quantify star clusters in terms of age, metallicity, and color
excess by their integrated photometry.  The well known
age-metallicity-extinction degeneracy was analyzed for various parameter
combinations, assuming different levels of photometric accuracy.  We
conclude that the {\it UBVRI} photometric system enables us to estimate
star cluster parameters over a wide range, if the overall photometric
accuracy is better than $\sim$\,0.03 mag.  \end{summary}


\resthead{Accuracy of star cluster parameters}
{D.~Narbutis, A.~Brid\v{z}ius, R.~Stonkut\.{e}, V.~Vansevi\v{c}ius}

\sectionb{1}{INTRODUCTION}

We investigate the possibility to quantify star cluster parameters (age
$t$, metallicity [M/H] and color excess $E_{B-V}$) comparing their
integrated color indices $U\!-\!V$, $B\!-\!V$, $V\!-\!R$ and $R\!-\!I$
with the simple stellar population (SSP) models.  The purpose of this
study is to evaluate the reliability of the derived parameters assuming
different levels of photometric accuracy, and to identify parameter
ranges susceptible to degeneracies.

For similar purposes various $\chi^{2}$ minimization techniques have
been used (e.g., Bik et al. 2003, Ma\'{i}z-Apell\'{a}niz 2004, de Grijs
et al. 2005).  The systematic uncertainties of quantified parameters,
associated with the selection of photometric passbands, have been
studied in detail by Anders et al.  (2004).  The importance of
interstellar extinction and star cluster metallicity has been widely
discussed based on various SSP models and parameter quantification
techniques by de Grijs et al.  (2005) . These studies concentrated on a
statistical analysis of the derived parameters for large samples of
artificial or real star clusters.

In this study we have estimated the {\it UBVRI} photometric system
capabilities to quantify star cluster parameters (age, metallicity and
color excess) and analyzed their degeneracies at various photometry
accuracy levels.

\sectionb{2}{THE METHOD}

We employed the P\'{E}GASE (v.~2.0; Fioc and Rocca-Volmerange 1997)
program package to compute SSP models ($U\!-\!V$, $B\!-\!V$, $V\!-\!R$
and $V\!-\!I$ color indices) of various ages and metallicities.  The
default P\'{E}GASE parameters were applied, but the universal initial
mass function (Kroupa 2002) was assumed.  The reddened color indices
were calculated as a function of color excess $E_{B-V}$, assuming the
standard extinction law (Cardelli, Clayton \& Mathis 1989).  The
three-parameter SSP model grid was constructed at the following nodes:
(1) 76 age ($t$) values from 1 Myr to 20 Gyr, with a constant step of
log\,($t/{\rm Myr}$) = 0.05, starting from $t=6$ Myr (for younger ages
grid nodes are located at 1, 2, 3, 4, 5 Myr); (2) 31 metallicity [M/H]
values from --2.3 to +0.7, with a step of 0.1 dex; (3) 201 color excess
$E_{B-V}$ values from 0.0 to 2.0, with a step of 0.01 (473\,556 models
in total).

The sample of 54 SSP models was selected from the model grid for further
analysis as the test star clusters ($t$ = 20, 50, 100, 200, 500 Myr and
1, 2, 5, 10 Gyr; [M/H]~=~0.0, --0.7, --1.7; $E_{B-V}$ = 0.1, 1.0).  The
first color excess value was chosen to represent a typical foreground
Milky Way extinction, inherent to extragalactic objects, and the second
-- substantially obscured star clusters in the spiral galaxy disks.

A star cluster parameter quantification code, using a technique similar
to the one developed for star quantification (Vansevi\v{c}ius \&
Brid\v{z}ius 1994), was implemented in the environment of data analysis
and graphing software package ``Origin''\footnote{~``OriginLab
Corporation''.}.  The parameter quantification is based on a comparison
of the observed color indices of star clusters with those of the SSP
models from the model grid.  For this purpose we use the quantification
criterion $\delta$, which is calculated according to the formula:
\begin{equation}
{\delta = \sqrt{ \sum_{i=1}^{n} W_{i} (CI_{i}^{\rm obs} -
CI_{i}^{\rm mod})^{2} \over \sum_{i=1}^{n} W_{i} }},
\end{equation}
where $CI_{i}^{\rm obs}$ stands for the observed star cluster color
indices $U\!-\!V$, $B\!-\!V$, $V\!-\!R$ and $V\!-\!I$; $CI_{i}^{\rm
mod}$ -- the corresponding color indices of SSP models from the grid;
$n$ -- the number of color indices used; $W _{i}$ -- the weights of the
observed color indices.  In the present study $CI_{i}^{\rm obs}$
represents color indices ($n=4$) of the test star clusters, and weights
assigned to color indices are equal ($W_{i}=1$).  The quantification
criterion $\delta$ mimics the effect of photometric accuracy.  In
general, it represents average observation errors of color indices.
Therefore, we use $\delta$ to study star cluster parameter degeneracies
and, by setting different $\delta_{\rm max}$ thresholds, determine
parameters and estimate their accuracy for a corresponding photometric
error budget.

The test star cluster colors are compared with the colors of all SSP
models from the model grid and $\delta$ is calculated at each grid node.
Then the test star cluster parameters ($t$, [M/H] and $E_{B-V}$) are
calculated as the SSP model parameter weighted averages, using only
those models, which have $\delta$ lower than the applied threshold
($\delta \leq \delta_{\rm max}$).  We use five $\delta_{\rm max}$ values
(0.01--0.05 mag, with a step of 0.01 mag) and provide the quantification
results as well as their standard deviations in Figures 1 and 2 for
color excess values $E_{B-V}=0.1$ and 1.0, respectively.  The weights
for parameter averaging and calculation of standard deviations were
assigned to 1 and to $(10^{-4}/\delta^{2})$ for the SSP models with
$\delta\leq0.01$ and $\delta>0.01$, respectively.

The well known age-metallicity-extinction degeneracy makes the procedure
described above not a trivial task, because of a few possible isolated
$\delta \leq \delta_{\rm max}$ ``islands'' in the parameter space.  To
overcome this problem we have chosen to select interactively a proper
``island'' for calculations of the parameters and their standard
deviations.  In reality such an improvement of the blind quantification
procedure could be based on multi-band star cluster environment images
or other {\it a priori} information, e.g., metallicity estimated by
spectroscopic methods.  Therefore, to calculate the weighted averages of
parameters and their standard deviations we used only the SSP models,
which reside in a single selected ``island'', satisfying the $\delta
\leq \delta_{\rm max}$ criterion.  This procedure excludes the models
located in secondary $\delta$ minima, which arise due to parameter
degeneracies.  For purposes of the present study, boundaries of a
continuous $\delta \leq \delta_{\rm max}$ ``island'' were determined by
starting the search from the global $\delta$ minimum, i.e., from the
position of the test star cluster under consideration in the SSP model
grid.  The parameter quantification maps for some characteristic test
star clusters are shown in Figures 3--8.

\sectionb{3}{RESULTS}

The summary of the parameter quantification results of the 54 test star
clusters is provided in Figures 1 and 2. In each panel the differences
of parameters (determined minus true) are shown in groups of five filled
circles (centered at their true age on $t$-axis), corresponding to the
quantification criterion values of $\delta \leq 0.01-0.05$ mag, with a
step of 0.01 mag (plotted from left to right).  Error bars indicate
standard deviations of the determined parameters and characterize sizes
of the $\delta \leq \delta_{\rm max}$ ``islands'' in the parameter
space.  Note that higher metallicity (Figure 1) and larger extinction
(Figure 2) make quantification procedure less accurate.

The parameter quantification maps of the representative star clusters of
age $t$ = 50 Myr, 500 Myr and 5 Gyr are displayed in Figures 3--8;
different shades of gray represent $\delta \leq 0.01$, 0.03 and 0.05 mag
(from the darkest to the lightest).  The analysis of parameter
quantification map patterns reveals, that the test star clusters can be
subdivided into three broad age groups according to the shape of the
age-metallicity degeneracy:  (1) younger than $\sim$\,100 Myr (Figures 3
and 4); (2) older than $\sim$\,100 Myr and younger than $\sim$\,1 Gyr
(Figures 5--6); (3) older than $\sim$\,1 Gyr (Figures 7--8).  Note,
however, that the age-extinction maps are much more regular, than the
age-metallicity maps, and the shape of $\delta$ ``islands'' gradually
changes from the youngest to the oldest cluster ages.  In general, the
parameter degeneracies seen in Figures 3--8 are mainly preconditioned by
the properties of the {\it UBVRI} photometric system itself, and should
not depend strongly on the P\'{E}GASE SSP models used for the
construction of the model grid.

The ultraviolet (UV) and infrared ({\it JHK}) passbands are known to be
helpful for breaking the age-metallicity degeneracy, see, e.g., de Grijs
et al.  (2003).  However, the accuracy of the UV and {\it JHK}
photometry is usually lower than the accuracy, which can be achieved in
{\it UBVRI} bands.  Therefore, the information on metallicity -- at
least a rough estimate from spectroscopy -- is of high importance for
improving the accuracy of the quantification procedure.  Even a
``realistic'' assumption, used in various star cluster studies (e.g., a
study of the M\,51 galaxy by Bik et al. 2003), helps to constrain
cluster ages.

We conclude that the {\it UBVRI} photometric system enables us to
estimate star cluster parameters over a wide range, if the overall
accuracy of color indices is better than $\sim$\,0.03 mag.

\vskip3mm
\enlargethispage{5mm}

ACKNOWLEDGMENTS. This work was financially supported in part by a Grant
of the Lithuanian State Science and Studies Foundation.

\begin{landscape}
\begin{figure}[!t]
\vbox{\centerline{\psfig{figure=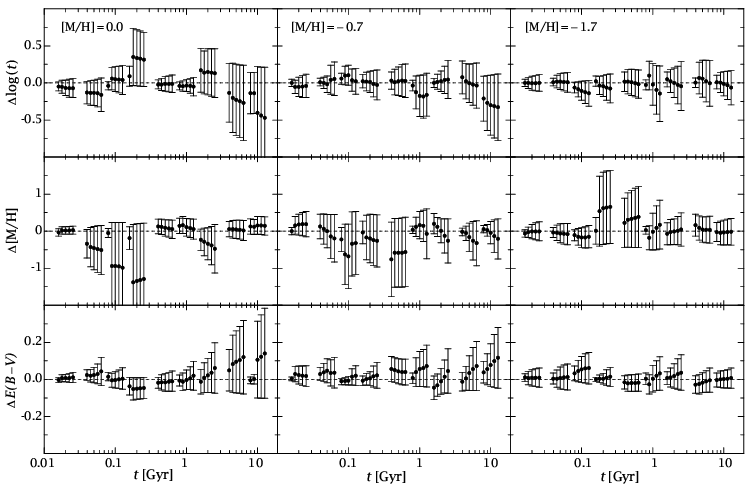,width=170truemm,angle=0,clip=}}
\vspace{-2mm}
\captionb{1}{Summarized test star cluster parameter quantification
results for the case of color excess $E_{B-V}$ = 0.1.  Different
metallicity ([M/H]~= 0.0, --0.7, --1.7) cases are shown in the left,
middle and right panels, respectively.  Differences of the parameters
(determined minus true) are shown in groups of five filled circles,
centered at the true test star cluster ages; within a group from left to
right the circles correspond to the quantification results derived at
$\delta_{\rm max}$ = 0.01--0.05 mag, with a step of 0.01 mag.  Error
bars indicate standard deviations of the derived parameters.}}
\end{figure}

\begin{figure}[!t]
\vbox{\centerline{\psfig{figure=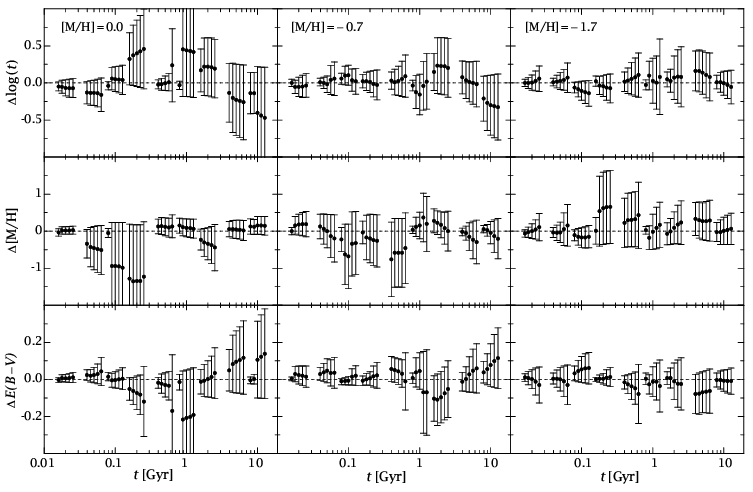,width=170truemm,angle=0,clip=}}
\captionc{2}{The same as in Figure 1, but for the case of
color excess $E_{B-V}$ = 1.0.}}
\end{figure}
\end{landscape}

\vbox{\centerline{\psfig{figure=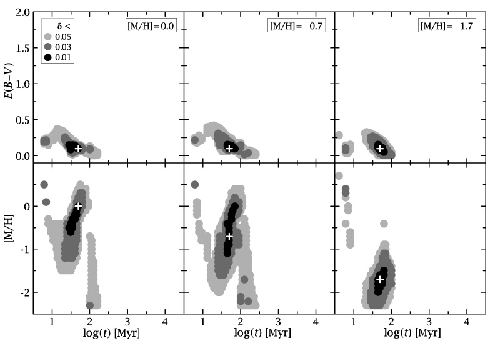,width=120truemm,angle=0,clip=}}
\vspace{-1mm}
\captionb{3}{Parameter quantification maps of the test star clusters of
age $t=50$ Myr, color excess $E_{B-V}$ = 0.1 and metallicity
[M/H]~=~0.0,~--0.7,~--1.7 (left, middle and right panels).  True
parameter values are marked with a white `plus' symbol.  Gray
levels correspond to a quantification criterion of
$\delta$\,$\leq$\,0.01, 0.03, 0.05 mag.}}

\vskip5mm
\vbox{\centerline{\psfig{figure=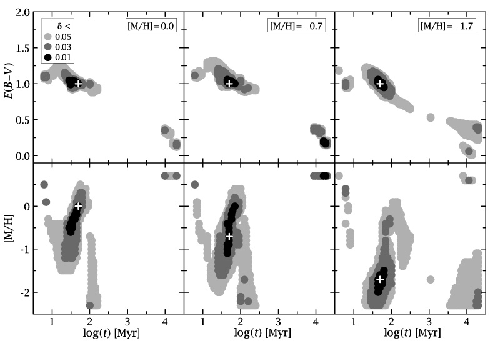,width=120truemm,angle=0,clip=}}
\vspace{-3mm}
\captionc{4}{The same as in Figure 3, but for the case of color excess
$E_{B-V}$ = 1.0.}}

\newpage
\vbox{\centerline{\psfig{figure=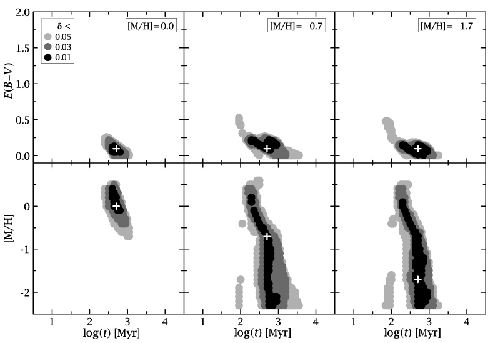,width=120truemm,angle=0,clip=}}
\vspace{-3mm}
\captionc{5}{The same as in Figure 3, but for the case of age $t=500$
Myr.}}

\vskip5mm

\vbox{\centerline{\psfig{figure=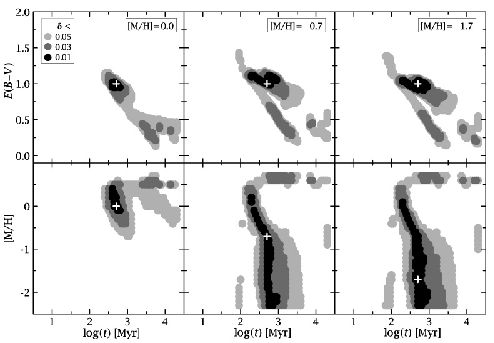,width=120truemm,angle=0,clip=}}
\vspace{-3mm}
\captionc{6}{The same as in Figure 4, but for the case of age $t=500$
Myr.}}

\newpage
\vbox{\centerline{\psfig{figure=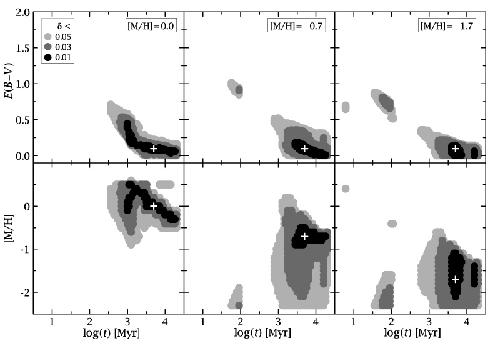,width=120truemm,angle=0,clip=}}
\vspace{-3mm}
\captionc{7}{The same as in Figure 3, but for the case of age $t=5$
Gyr.}}
\vskip5mm
\vbox{\centerline{\psfig{figure=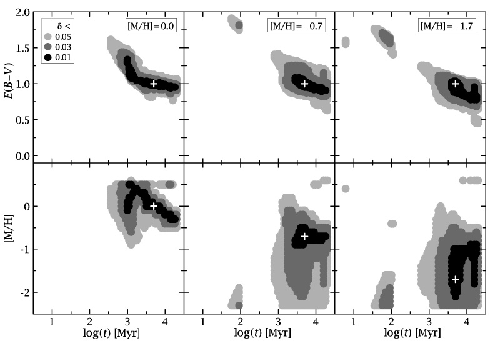,width=120truemm,angle=0,clip=}}
\vspace{-3mm}
\captionc{8}{The same as in Figure 4, but for the case of age $t=5$
Gyr.}}
\newpage

\References

\refb Anders~P., Bissantz~N., Fritze-v.~Alvensleben~U., de~Grijs~R.
2004, MNRAS, 347, 196

\refb Bik~A., Lamers~H.\,J.\,G.\,L.\,M., Bastian~N., Panagia~N.,
Romaniello~M. 2003, A\&A, 397, 473

\refb Cardelli~J.~A., Clayton~G.~C., Mathis~J.~S. 1989, ApJ, 345, 245

\refb Fioc~M., Rocca-Volmerange~B. 1997, A\&A, 326, 950

\refb de Grijs~R., Anders~P., Lamers~H.\,J.\,G.\,L.\,M., Bastian~N.,
Fritze-v.  Alvensleben~U., Parmentier~G., Sharina~M.~E., Yi~S. 2005,
MNRAS, 359, 874

\refb de Grijs~R., Fritze-v.  Alvensleben~U., Anders~P., Gallagher
III~J.~S., Bastian~N., Taylor~V.~A., Windhorst~R.~A. 2003, MNRAS, 342,
259

\refb Kroupa~P. 2002, Science, 295, 82

\refb Ma\'{i}z-Apell\'{a}niz J. 2004, PASP, 116, 859

\refb Vansevi\v{c}ius V., Brid\v{z}ius A. 1994, Baltic Astronomy, 3, 193

\end{document}